\documentclass[aps,pra,twocolumn,showpacs]{revtex4}
\usepackage{amsmath}
\usepackage{amsfonts}
\usepackage{amssymb}
\usepackage{graphicx}

\begin{document}
\title{Asymmetric quantum hypothesis testing with Gaussian states}
\author{Gaetana Spedalieri}
\affiliation{Department of Computer Science, University of York, York YO10 5GH, United Kingdom}
\author{Samuel L. Braunstein}
\affiliation{Department of Computer Science, University of York, York YO10 5GH, United Kingdom}

\begin{abstract}
We consider the asymmetric formulation of quantum hypothesis testing, where
two quantum hypotheses have different associated costs. In this problem, the
aim is to minimize the probability of false negatives and the optimal
performance is provided by the quantum Hoeffding bound. After a brief review
of these notions, we show how this bound can be simplified for pure states. We
then provide a general recipe for its computation in the case of multimode
Gaussian states, also showing its connection with other easier-to-compute
lower bounds. In particular, we provide analytical formulae and numerical
results for important classes of one- and two-mode Gaussian states.

\end{abstract}

\pacs{03.67.-a, 89.70.Cf, 03.67.Hk, 03.65.Ta, 02.10.Ud}
\maketitle

\section{Introduction}

Quantum hypothesis testing (QHT) is a fundamental topic in quantum information
theory~\cite{Wilde,Nielsen}, playing a non-trivial role in protocols of
quantum communication and quantum cryptography~\cite{QCprot,QCprot2}. The
typical formulation of QHT is given in terms of quantum state
discrimination~\cite{QHT,QHT2,RMP,Helstrom}, where a certain number of
generally non-orthogonal quantum states (the quantum hypotheses) have to be
discriminated by means of a quantum measurement. In particular, the simplest
scenario regards the statistical discrimination between two non-orthogonal
quantum states, corresponding to the `null' and the `alternative' quantum
hypotheses, occurring with some a priori probabilities. In \textit{symmetric
testing}, these hypotheses have the same cost~\cite{QHT,QHT2,Helstrom} and the
goal is to minimize the mean error probability of confusing them by suitably
optimizing the quantum measurement.

For such a basic problem, we know closed analytical formulae identifying both
the minimum error probability, given by the Helstrom bound~\cite{Helstrom},
and the optimal quantum detection, expressed in terms of the Helstrom
matrix~\cite{Helstrom}. Furthermore, we can also use an easier-to-compute
bound which becomes tight in asymptotic conditions. This is the
recently-introduced quantum Chernoff bound~\cite{QCB}, for which we know
simple formulae in the case of multi-mode Gaussian states~\cite{QCB2}, (i.e.,
those states with Gaussian Wigner function~\cite{RMP}).

In this paper, we consider \textit{asymmetric} QHT, where two quantum
hypotheses have different associated costs~\cite{QHT,QHT2,Helstrom}. In this
approach, we aim to minimize the probability that the alternative hypothesis
is confused for the null hypothesis, an error which is known as `false
negative'. This minimization has to be done by suitably constraining the
probability of another possible error, known as a `false positive', where the
null hypothesis is confused for the alternative hypothesis. This is clearly
the best approach for instance in medical-type testing, where the null
hypothesis typically represents absence of a disease, while the alternative
corresponds to the presence of a disease.

Asymmetric QHT is typically formulated as a multi-copy discrimination problem,
where a large number of copies of the two possible states are prepared and
subjected to a collective quantum measurement. From this point of view, the
aim is to maximize the error-exponent describing the exponential decay of the
false negatives, while placing a reasonable constraint on the false positives.
For this calculation, we can rely on two mathematical tools. The first is the
quantum relative entropy~\cite{RMP} between the two states, while the other is
the recently-introduced quantum Hoeffding bound (QHB)~\cite{QHB}, which
performs the optimization of the error-exponent while providing a better
control on the false positives.

In this work, we start by giving some basic notions on asymmetric QHT and
briefly reviewing the QHB, also showing how its computation simply reduces to
the quantum fidelity~\cite{Fidelity} in the presence of pure states. Then, we
provide a general recipe for computing this bound in the case of multimode
Gaussian states, for which\ it can be expressed in terms of their first- and
second-order statistical moments. In the general multimode case, we derive a
relation between the QHB and other easier-to-compute bounds, which are based
on well-known mathematical inequalities. Finally, we derive analytical
formulas and numerical results for the most important classes of one-mode and
two-mode Gaussian states.

By developing the theory of asymmetric QHT for Gaussian states, our work could
be useful in tasks and protocols involving Gaussian quantum
information~\cite{RMP}, including technological applications of quantum
channel discrimination (e.g., quantum illumination~\cite{Qill,Qill2} or
quantum reading~\cite{Qread,Qread2,Qread3,Qread4}) where we are interested in
increasing our ability to accept one specific quantum hypothesis.

\section{Brief review of asymmetric testing}

\subsection{Basic formulation}

In binary QHT we consider a quantum system which is prepared in some unknown
quantum state $\rho$, which can be $\rho_{0}$\ or $\rho_{1}$. For instance we
can imagine one party, say Alice, who prepares such a system. This system is
then passed to Bob, who does not know which choice Alice has made. Thus, Bob
must decide between the following two hypotheses
\begin{align}
\text{Null hypothesis}~H_{0}  &  :\rho=\rho_{0}~,\\
\text{Alternative hypothesis~}H_{1}  &  :\rho=\rho_{1}\text{~.}%
\end{align}
In order to discriminate between these two hypotheses, i.e., distinguish
between the two states, Bob applies a quantum measurement, generally described
by a positive operator valued measure (POVM). Without loss of generality, Bob
can always reduce his measurement to be a dichotomic POVM $\left\{  \Pi
_{k}\right\}  $ with $k=0,1$~\cite{Helstrom}. The outcome $k=0$, with POVM
operator $\Pi_{0}$, is associated to the null hypotheses $H_{0}$, while the
other outcome $k=1$, with POVM operator $\Pi_{1}=I-\Pi_{0}$, is associated
with the alternative hypothesis $H_{1}$.

Since the two quantum states $\rho_{0}$\ and $\rho_{1}$ are generally
non-orthogonal, there is a non-zero error probability to confuse the two
hypotheses. We can identify two different types of error: Type-I and type-II
errors, with associated conditional error probabilities. By definition, the
type-I error, also known as a `false-positive', is where Bob accepts the
alternative hypothesis $H_{1}$ when the null hypothesis $H_{0}$\ holds. We
have a corresponding error probability expressed by
\begin{equation}
\alpha:=p(H_{1}|H_{0})=\mathrm{Tr}(\Pi_{1}\rho_{0}).
\end{equation}
Then, the type-II error or `false-negative' is where Bob accepts the null
hypothesis $H_{0}$ when the true hypothesis is the alternative$\ H_{1}$. This
error occurs with conditional probability%
\begin{equation}
\beta:=p(H_{0}|H_{1})=\mathrm{Tr}(\Pi_{0}\rho_{1}).
\end{equation}

Note that we can introduce other probabilities, but they are fully determined
by $\alpha$ and $\beta$. For instance, we may also consider the `specificity'
or `true-negativity' of the test which is the success probability of
identifying the null hypothesis, i.e., $p(H_{0}|H_{0})$ which is simply given
by $1-\alpha$. Similarly, we may also consider the `sensitivity' or
`true-positivity' of the test which is the success probability of identifying
the alternative hypothesis, i.e., $p(H_{1}|H_{1})=1-\beta$.

The costs associated with the two types of error can be very different
especially in the medical and histological settings. For instance, in a
medical test, $H_{0}$\ is typically associated with no illness, while $H_{1}%
$\ with the presence of the disease. It is therefore clear that we would like
to have tests where the false-negative probability (or rate) $\beta$\ is the
lowest possible, so that ill patients are not diagnosed as healthy. For this
reason, in a medical setting, hypothesis testing is almost always asymmetric,
meaning that we aim to minimize one of the two conditional error probabilities.

\subsection{Multi-copy formulation}

In general we can formulate the problem of QHT as an $M$-copy discrimination
problem~\cite{QHT,QHT2}. This means that Alice has $M$ quantum systems which
are prepared in two possible multi-copy states%
\begin{align}
H_{0}  &  :\rho=\rho_{0}^{\otimes M}=\rho_{0}\otimes...\otimes\rho_{0}~,\\
H_{1}  &  :\rho=\rho_{1}^{\otimes M}=\rho_{1}\otimes...\otimes\rho
_{1}~.\nonumber
\end{align}
These systems are passed to Bob who performs a collective measurement on them.
As before, this general POVM can be chosen to be dichotomic $\left\{  \Pi
_{0},\Pi_{1}\right\}  $ with $\Pi_{1}=I-\Pi_{0}$.

The error probabilities now depend on the number of copies $M$. In particular,
the probability of false positives is given by%
\begin{equation}
\alpha_{M}:=p(H_{1}|H_{0})=\mathrm{Tr}(\Pi_{1}\rho_{0}^{\otimes M}),
\end{equation}
and the probability of false negatives is%
\begin{equation}
\beta_{M}:=p(H_{0}|H_{1})=\mathrm{Tr}(\Pi_{0}\rho_{1}^{\otimes M}).
\end{equation}

In the limit of a large number of copies $(M\gg1)$, these probabilities go to
zero exponentially, i.e., we have%
\begin{equation}
\alpha_{M}\simeq\frac{1}{2}e^{-\alpha_{R}M},~\beta_{M}\simeq\frac{1}%
{2}e^{-\beta_{R}M},
\end{equation}
where the coefficients
\begin{align}
\alpha_{R}  &  =-\lim_{M\rightarrow+\infty}\frac{1}{M}\ln\alpha_{M}~,\\
\beta_{R}  &  =-\lim_{M\rightarrow+\infty}\frac{1}{M}\ln\beta_{M}~,
\end{align}
are called the `error-exponents' or `rate limits'~\cite{QHB}.

Bob's aim is to\ maximize the error exponent $\beta_{R}$, so that the error
probability of false negatives $\beta_{M}$ has the fastest exponential decay
to zero. This must be done while controlling the rate of false positives. Here
a well known result is the `quantum Stein lemma'~\cite{QHB} which connects
$\beta_{R}$\ with the quantum relative entropy between the single-copy states
$\rho_{0}$\ and $\rho_{1}$. For a large number of copies $M\gg1$, there is a
dichotomic POVM such that the error probability of the false positives is
bounded
\begin{equation}
\alpha_{M}\leq\varepsilon\text{~~for any~}0<\varepsilon<1,
\end{equation}
and the error probability of false negatives goes to zero with error-exponent%
\begin{equation}
\beta_{R}=S(\rho_{0}||\rho_{1})=\mathrm{Tr}\rho_{0}(\ln\rho_{0}-\ln\rho_{1}).
\label{12}%
\end{equation}

More powerfully, we may use the notion of the QHB~\cite{QHB}. For $M\gg1$,
there is a dichotomic POVM such that the error-exponent of false positives is
lower-bounded by a positive parameter%
\begin{equation}
\alpha_{R}\geq r\text{ for any }r>0,
\end{equation}
and the error-exponent of false negatives satisfies%
\begin{equation}
\beta_{R}=H(r),
\end{equation}
where $H(r)\geq0$ is the QHB defined by%
\begin{equation}
H(r):=\sup_{0\leq s<1}P(r,s),~~P(r,s):=\frac{-r~s-\ln C_{s}}{1-s},
\label{QHB2}%
\end{equation}
where
\begin{equation}
C_{s}:=\mathrm{Tr}(\rho_{0}^{s}\rho_{1}^{1-s}) \label{sOVER}%
\end{equation}
is the `s-overlap' between the single-copy states $\rho_{0}$ and $\rho_{1}$.
Note that the quantum Hoeffding bound enforces a stronger constraint on
false-positives, since these are bounded at the level of the error-exponent
and not at the level of the error probability as happens for the quantum
relative entropy bound.

\section{Asymmetric testing with pure states}

Asymmetric testing becomes very simple when one of the states (or both) is
pure. In this case, we can in fact relate the QHB to the quantum fidelity
between the two states.

Let us start by considering the case where only one of the states is pure,
e.g., $\rho_{0}=\left\vert \psi_{0}\right\rangle \left\langle \psi
_{0}\right\vert $. We can write~\cite{JPAgae}%
\begin{equation}
\inf_{s}C_{s}=F(\left\vert \psi_{0}\right\rangle ,\rho_{1}), \label{relFCs}%
\end{equation}
where $F$ is the fidelity between $\left\vert \psi_{0}\right\rangle $ and
$\rho_{1}$. Eq.~(\ref{relFCs}) implies $C_{s}\geq F$. By using the latter
inequality in Eq.~(\ref{QHB2}), we derive the fidelity-bound%
\begin{equation}
H(r)\leq H_{F}(r):=\sup_{0\leq s<1}\frac{-r~s-\ln F}{1-s}~. \label{FQHB}%
\end{equation}

This bound can be further simplified by explicitly performing the maximization
with regard to the parameter $s$. After a simple calculation we find%
\begin{equation}
H_{F}(r)=\left\{
\begin{array}
[c]{c}%
\ln\frac{1}{F},\text{~~~~for~}r\geq\ln\frac{1}{F}~,\\
\\
+\infty,\text{~~~~for~}r<\ln\frac{1}{F}~,
\end{array}
\right.  \label{QHBF2}%
\end{equation}
which depends on the comparison between the parameter $r$ and the fidelity $F$
of the two states.

More specifically, in the discrimination of two pure states, we find that the
previous fidelity-bound becomes tight
\begin{equation}
H(r)=H_{F}(r)~. \label{Fidtight}%
\end{equation}
In fact, for pure states $\rho_{0}=|\psi_{0}\rangle\langle\psi_{0}|$ and
$\rho_{1}=|\psi_{1}\rangle\langle\psi_{1}|$, and for any $0<s<1$, we can write%
\begin{align}
C_{s}  &  =\mathrm{Tr}(|\psi_{0}\rangle\langle\psi_{0}|^{s}|\psi_{1}%
\rangle\langle\psi_{1}|^{1-s})=\mathrm{Tr}(|\psi_{0}\rangle\langle\psi
_{0}|\psi_{1}\rangle\langle\psi_{1}|)\nonumber\\
&  =|\langle\psi_{0}|\psi_{1}\rangle|^{2}=F(|\psi_{0}\rangle,|\psi_{1}%
\rangle).
\end{align}
Therefore we can replace $\ln C_{s}=\ln F$ in the QHB of Eq.~(\ref{QHB2}),
which implies Eq.~(\ref{Fidtight})~\cite{NoteGG}.

\section{Asymmetric testing with Gaussian states\label{Sec_AsyGAUSS}}

\subsection{Basics of bosonic systems and Gaussian states}

A bosonic system of $n$ modes is a quantum system described by a tensor
product Hilbert space $\mathcal{H}^{\otimes n}$ and a vector of quadrature
operators~\cite{BraR1,BraR2}
\begin{equation}
\mathbf{\hat{x}}^{T}:=(\hat{q}_{1},\hat{p}_{1},\ldots,\hat{q}_{n},\hat{p}%
_{n}).
\end{equation}
These operators satisfy the vectorial commutation relations~\cite{SPEDcomm}%
\begin{equation}
\lbrack\mathbf{\hat{x}},\mathbf{\hat{x}}^{T}]:=\mathbf{\hat{x}\hat{x}}%
^{T}-(\mathbf{\hat{x}\hat{x}}^{T})^{T}=2i\mathbf{\Omega}~, \label{CommQUAD}%
\end{equation}
where $\mathbf{\Omega}$ is the symplectic form, defined as%
\begin{equation}
\mathbf{\Omega}:=\bigoplus\limits_{k=1}^{n}\left(
\begin{array}
[c]{cc}%
0 & 1\\
-1 & 0
\end{array}
\right)  ~. \label{Symplectic_Form}%
\end{equation}
Correspondingly, a real matrix $\mathbf{S}$ is called `symplectic'\ when it
preserves $\mathbf{\Omega}$ by congruence, i.e., $\mathbf{S\Omega S}%
^{T}=\mathbf{\Omega}$.

By definition, we say that a bosonic state $\rho$ is `Gaussian'\ when its
phase-space Wigner representation is Gaussian~\cite{RMP}. In such a case, we
can completely describe the state by means of its first- and second-order
statistical moments. These are the mean value or displacement vector
$\mathbf{\bar{x}}:=\mathrm{Tr}(\mathbf{\hat{x}}\rho)$, and the covariance
matrix (CM) $\mathbf{V}$ with generic element%
\begin{equation}
V_{ij}=\tfrac{1}{2}\mathrm{Tr}(\{\hat{x}_{i},\hat{x}_{j}\}\rho)-\bar{x}%
_{i}\bar{x}_{j}~,
\end{equation}
where $\{,\}$ denotes the anticommutator. The CM\ is a $2n\times2n$ real
symmetric matrix, which must satisfy the uncertainty principle~\cite{RMP}%
\begin{equation}
\mathbf{V}+i\mathbf{\Omega}\geq0~. \label{unc_PRINC}%
\end{equation}

An important tool in the manipulation of Gaussian states is Williamson's
theorem~\cite{RMP}: For any CM\ $\mathbf{V}$, there is a symplectic matrix
$\mathbf{S}$ such that%
\begin{equation}
\mathbf{V}=\mathbf{SWS}^{T}~,
\end{equation}
where%
\begin{equation}
\mathbf{W}=\bigoplus\limits_{k=1}^{n}\nu_{k}\mathbf{I}\text{~},~\mathbf{I}%
:=\left(
\begin{array}
[c]{cc}%
1 & 0\\
0 & 1
\end{array}
\right)  .
\end{equation}
The matrix $\mathbf{W}$\ is the `Williamson form'\ of $\mathbf{V}$, and the
set $\{\nu_{1},\cdots,\nu_{n}\}$ is the `symplectic spectrum'\ of $\mathbf{V}%
$. According to the uncertainty principle, each symplectic eigenvalue must
satisfy the condition $\nu_{k}\geq1$, with $\nu_{k}=1$ for all $k$ if and only
if the Gaussian state is pure.

\subsection{Computation of the quantum Hoeffding bound}

Our goal is to find a general recipe for the calculation of the QHB for
Gaussian states. We start from the general formula in Eq.~(\ref{QHB2})
involving the logarithm of the $s$-overlap $C_{s}$ defined in Eq.~(\ref{sOVER}%
). Given two $n$-mode Gaussian states, $\rho_{0}$ and $\rho_{1}$, we can write
an explicit Gaussian formula for the $s$-overlap in terms of their statistical
moments ($\mathbf{\bar{x}}_{0}$, $\mathbf{V}_{0}$) and ($\mathbf{\bar{x}}_{1}%
$, $\mathbf{V}_{0}$). This is given by~\cite{QCB2,JPAgae}%
\begin{equation}
C_{s}=\frac{\Pi_{s}}{\sqrt{\det\boldsymbol{\Sigma}_{s}}}\exp\left[
-\frac{\mathbf{d}^{T}\boldsymbol{\Sigma}_{s}^{-1}\mathbf{d}}{2}\right]  ,
\label{sOVERLAP}%
\end{equation}
where $\mathbf{d}:=\mathbf{\bar{x}}_{0}-\mathbf{\bar{x}}_{1}$ is the
difference between the mean values, while $\Pi_{s}$ and $\boldsymbol{\Sigma
}_{s}$ depends on the CMs $\mathbf{V}_{0}$ and $\mathbf{V}_{1}$. In
particular, introducing the two real functions
\begin{align}
G_{s}(x)  &  :=\frac{2^{s}}{(x+1)^{s}-(x-1)^{s}},\label{Gfunc}\\
\Lambda_{s}(x)  &  :=\frac{(x+1)^{s}+(x-1)^{s}}{(x+1)^{s}-(x-1)^{s}},
\label{Lfunc}%
\end{align}
we can write the formulas%
\begin{equation}
\Pi_{s}:=2^{n}\Pi_{k=1}^{n}G_{s}(\nu_{k}^{0})G_{1-s}(\nu_{k}^{1})~,
\end{equation}
and%
\begin{align}
\boldsymbol{\Sigma}_{s}  &  :=\mathbf{S}_{0}\mathbf{~}[\oplus_{k=1}^{n}%
\Lambda_{s}(\nu_{k}^{0})\mathbf{I}]~\mathbf{S}_{0}^{T}\nonumber\\
&  +\mathbf{S}_{1}\mathbf{~}[\oplus_{k=1}^{n}\Lambda_{1-s}(\nu_{k}%
^{1})\mathbf{I}]~\mathbf{S}_{1}^{T}, \label{SIGMA}%
\end{align}
where $\{\nu_{k}^{0}\}$ and $\{\nu_{k}^{1}\}$ are the symplectic spectra of
the two states, with $\mathbf{S}_{0}$ and $\mathbf{S}_{1}$ being the
symplectic matrices which diagonalize the two CMs according to Williamson's
theorem, i.e.,%
\begin{equation}
\mathbf{V}_{0}=\mathbf{S}_{0}\mathbf{~}(\oplus_{k=1}^{n}\nu_{k}^{0}%
\mathbf{I})~\mathbf{S}_{0}^{T},~\mathbf{V}_{1}=\mathbf{S}_{1}\mathbf{~}%
(\oplus_{k=1}^{n}\nu_{k}^{1}\mathbf{I})~\mathbf{S}_{1}^{T}.
\end{equation}

Substituting Eq.~(\ref{sOVERLAP}) into Eq.~(\ref{QHB2}), corresponds to
explicitly computing the logarithmic term $\ln C_{s}$, yielding%
\begin{equation}
\ln C_{s}=\ln\Pi_{s}-\frac{1}{2}\left\{  \ln\det\mathbf{\Sigma}_{s}%
+\mathbf{d}^{T}\boldsymbol{\Sigma}_{s}^{-1}\mathbf{d}\right\}  ~. \label{lnCs}%
\end{equation}
In particular for zero-mean Gaussian states we have $\mathbf{d}=0$ and the
previous expression simplifies to%
\begin{equation}
\ln C_{s}=\ln\Pi_{s}-\frac{1}{2}\ln\det\boldsymbol{\Sigma}_{s}~.
\end{equation}

\subsection{Other computable bounds}

Note that computing the $s$-overlap $C_{s}$ and its logarithmic form $\ln
C_{s}$\ could be difficult due to the presence of the symplectic matrices,
$\mathbf{S}_{0}$ and $\mathbf{S}_{1}$, in the term $\boldsymbol{\Sigma}_{s}$
in Eq.~(\ref{SIGMA}). A possible solution is to compute an upper bound, known
as the `Minkowski bound', which is based on the Minkowski determinant
inequality~\cite{Bathia} and depends only on the two symplectic
spectra~\cite{QCB2}. Specifically, we have $C_{s}\leq M_{s}$, where%
\begin{equation}
M_{s}:=4^{n}\left[  \prod\limits_{k=1}^{n}\Psi_{s}(\nu_{k}^{0},\nu_{k}%
^{1})+\prod\limits_{k=1}^{n}\Psi_{1-s}(\nu_{k}^{1},\nu_{k}^{0})\right]  ^{-n},
\label{Sum_bound}%
\end{equation}
and%
\begin{align}
\Psi_{s}(x,y)  &  :=\{[\left(  x+1\right)  ^{s}+\left(  x-1\right)
^{s}]\nonumber\\
&  \times\lbrack\left(  y+1\right)  ^{1-s}-\left(  y-1\right)  ^{1-s}%
]\}^{1/n}. \label{PSI}%
\end{align}
Another easy-to-compute upper bound is the `Young bound'\ $Y_{s}$, which is
based on Young's inequality~\cite{Young} and satisfies
\begin{equation}
C_{s}\leq M_{s}\leq Y_{s}, \label{boundsCOMP}%
\end{equation}
where~\cite{QCB2}%
\begin{equation}
Y_{s}:=2^{n}\prod\limits_{k=1}^{n}\Gamma_{s}(\nu_{k}^{0})\Gamma_{1-s}(\nu
_{k}^{1})~, \label{Product_bound}%
\end{equation}
and%
\begin{equation}
\Gamma_{s}(x):=\left[  (x+1)^{2s}-(x-1)^{2s}\right]  ^{-\frac{1}{2}}~.
\label{GAMMA}%
\end{equation}

Taking the negative logarithm of Eq.~(\ref{boundsCOMP}), we can write the
following inequality for the QHB%
\begin{equation}
H(r)\geq H_{M}(r)\geq H_{Y}(r),~
\end{equation}
where
\begin{align}
H_{M}(r)  &  :=\sup_{0\leq s<1}\frac{-r~s-\ln M_{s}}{1-s},\\
H_{Y}(r)  &  :=\sup_{0\leq s<1}\frac{-r~s-\ln Y_{s}}{1-s}.
\end{align}

In the specific case where one of the two Gaussian states is pure, we can
compute their fidelity $F$ and apply the upper bound given in Eqs.~(\ref{FQHB}%
) and~(\ref{QHBF2}), which becomes tight when both states are pure [see
Eq.~(\ref{Fidtight})]. In particular, for two multimode Gaussian states
$\rho_{0}=\left\vert \psi_{0}\right\rangle \left\langle \psi_{0}\right\vert $
and $\rho_{1}$, we can easily write their fidelity $F$ in terms of the
statistical moments~\cite{JPAgae}%
\begin{equation}
F=\frac{2^{n}}{\sqrt{\det\mathbf{L}}}\exp\left(  -\frac{\mathbf{d}%
^{T}\mathbf{L}^{-1}\mathbf{d}}{2}\right)  , \label{Fid_FormulaNEW}%
\end{equation}
where $\mathbf{L}:=\mathbf{V}_{0}+\mathbf{V}_{1}$. As a result, we can use
Eq.~(\ref{QHBF2}) with%
\begin{equation}
\ln\frac{1}{F}=\frac{1}{2}\left[  \ln\left(  \frac{\det\mathbf{L}}{4^{n}%
}\right)  +\mathbf{d}^{T}\mathbf{L}^{-1}\mathbf{d}\right]  .
\end{equation}

\section{Discrimination of one-mode Gaussian states}

In this section, we examine the case of one-mode Gaussian states. This means
we fix $n=1$ in the previous formulas of Sec.~\ref{Sec_AsyGAUSS}, with
matrices becoming $2\times2$, vectors becoming 2-dimensional, and symplectic
spectra reducing to a single eigenvalue. For instance, the $s$-overlap can be
more simply computed using the expressions%
\begin{align}
\Pi_{s}  &  =2~G_{s}(\nu^{0})~G_{1-s}(\nu^{1}),\label{PI1}\\
\boldsymbol{\Sigma}_{s}  &  =\Lambda_{s}(\nu^{0})~\mathbf{S}_{0}\mathbf{S}%
_{0}^{T}+\Lambda_{1-s}(\nu^{1})~\mathbf{S}_{1}\mathbf{S}_{1}^{T}.
\label{SIGMA1}%
\end{align}
In particular, here we shall derive the analytic formulas for the QHB for two
important classes: Coherent states (in Sec.~\ref{Sec_AMP}) and thermal states
(in Sec.~\ref{Sec_THERMAL}).

\subsection{Asymmetric testing of coherent amplitudes\label{Sec_AMP}}

The expression of the QHB\ is greatly simplified in the case of one-mode
coherent states $\rho_{0}=\left\vert \alpha_{0}\right\rangle \left\langle
\alpha_{0}\right\vert $ and $\rho_{1}=\left\vert \alpha_{1}\right\rangle
\left\langle \alpha_{1}\right\vert $. Since both states are pure, the QHB is
equal to the fidelity bound in Eq.~(\ref{QHBF2}), i.e., $H(r)=H_{F}(r)$.
Therefore, it is sufficient to compute the fidelity between the two coherent
states, which is given by
\begin{equation}
F=\left\vert \left\langle \alpha_{0}\right\vert \left.  \alpha_{1}%
\right\rangle \right\vert ^{2}=e^{-\left\vert \alpha_{0}-\alpha_{1}\right\vert
^{2}}, \label{FidCOH}%
\end{equation}
so that $\ln\frac{1}{F}=\left\vert \alpha_{0}-\alpha_{1}\right\vert
^{2}:=\sigma$, and we can write%
\begin{equation}
H(r)=\left\{
\begin{array}
[c]{c}%
\sigma\text{~,~~~~~~for~}r\geq\sigma~,\\
\\
+\infty\text{~,~~~for~}r<\sigma~.
\end{array}
\right.
\end{equation}

Assuming that we impose a good control on the rate of false positives (so that
$r\geq\sigma$), then the error-exponent for the false negatives is simply
given by $H(r)=\sigma$. More explicitly, this corresponds to an asymptotic
error rate%
\begin{equation}
\beta_{M}=\frac{1}{2}e^{-M\sigma}=\frac{F^{M}}{2}~.
\end{equation}

Note that, if we have poor control on the rate of false positives, i.e.,
$r<\sigma$, then the QHB $H(r)$ is infinite. This means that the probability
of false negatives $\beta_{M}$ goes to zero super-exponentially, i.e., more
quickly than any decreasing exponential function.

\subsection{Asymmetric testing of thermal noise\label{Sec_THERMAL}}

In this section we derive the QHB for one-mode thermal states $\rho_{0}%
=\rho_{\text{th}}(\nu^{0})$ and $\rho_{1}=\rho_{\text{th}}(\nu^{1})$, with
variances equal to $\nu^{0}$ and $\nu^{1}$, respectively (in our notation,
$\nu=2\bar{n}+1$, where $\bar{n}$ is the mean number of thermal photons).
These Gaussian states have zero mean ($\mathbf{\bar{x}}_{0}=\mathbf{\bar{x}%
}_{1}=0$) and CMs in the Williamson form $\mathbf{V}_{0}=\nu^{0}\mathbf{I}$
and $\mathbf{V}_{1}=\nu^{1}\mathbf{I}$ (so that $\mathbf{S}_{0}=\mathbf{S}%
_{1}=\mathbf{I}$). Thus, we can write%
\begin{equation}
\boldsymbol{\Sigma}_{s}=\varepsilon_{s}\mathbf{I},~\varepsilon_{s}%
:=\Lambda_{s}(\nu^{0})+\Lambda_{1-s}(\nu^{1}),
\end{equation}
and derive%
\begin{equation}
C_{s}=\frac{\Pi_{s}}{\varepsilon_{s}}=\frac{2}{(\nu^{0}+1)^{s}(\nu
^{1}+1)^{1-s}-(\nu^{0}-1)^{s}(\nu^{1}-1)^{1-s}}.
\end{equation}
This is the $s$-overlap to be used in the QHB of Eq.~(\ref{QHB2}).

\begin{figure}[tbh] \vspace{-1.0cm}
\par
\begin{center}
\includegraphics[width=0.50\textwidth] {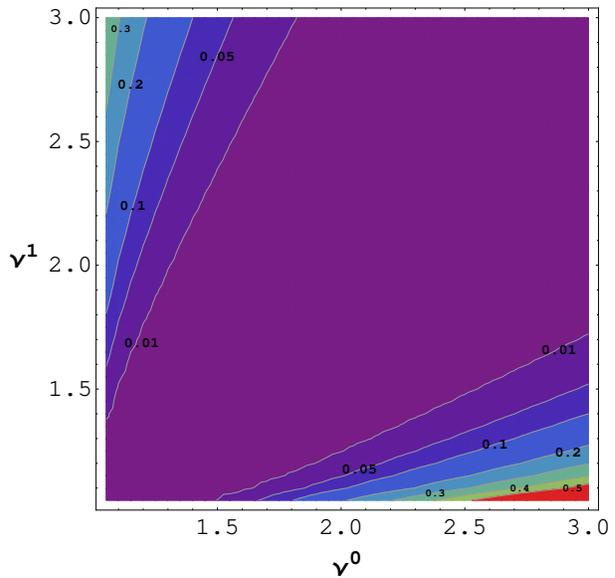}
\end{center}
\par
\vspace{-0.5cm}\caption{(Color online). We plot the QHB associated
with the discrimination of two thermal states:
$\rho_{\text{th}}(\nu^{0})$ as null hypothesis,\ and
$\rho_{\text{th}}(\nu^{1})$ as alternative hypothesis. We consider
low thermal variances $1<\nu^{0},\nu^{1}\leq3$ and we have set
$r=0.1$ for the false positives.}%
\label{PicThermal}%
\end{figure}

Given two arbitrary $\nu^{0}\geq1$ and $\nu^{1}\geq1$, the
maximization in Eq.~(\ref{QHB2}) can be done numerically. The
results are shown in Fig.~\ref{PicThermal} for thermal states with
variances up to $3$ vacuum units (equivalent to $1$ mean thermal
photon). From the figure we can see an asymmetry with respect to
the bisector $\nu^{0}=\nu^{1}$ which is a consequence of the
asymmetric nature of the hypothesis test. The bottom-right part of
the figure is related to the minimum probability of confusing a
nearly-vacuum state ($\nu^{1}\simeq1$) with a thermal state having
one average photon ($\nu^{0}\simeq3$). By contrast, the top-left
part of the figure is related to the probability of confusing a
thermal state having one average photon ($\nu^{1}\simeq3$) with a
nearly-vacuum state ($\nu^{0}\simeq1$). These probabilities are
clearly different.

We are able to derive a simple analytical result when we compare a thermal
state with the vacuum state. Let us start by considering the vacuum state to
be the null hypothesis ($\nu^{0}=1$) while the thermal state is the
alternative hypothesis ($\nu^{1}:=\nu>1$). In this specific case, we find%
\begin{equation}
\ln C_{s}=(1-s)\ln\left(  \frac{2}{1+\nu}\right)  ~,
\end{equation}
and we get%
\begin{equation}
P(r,s)=\ln\left(  \frac{1+\nu}{2}\right)  -\frac{rs}{1-s}~.
\end{equation}

Since $\nu$ is a constant, the maximization of $P$ over $0\leq s<1$
corresponds to minimizing the function $rs(1-s)^{-1}$, whose minimum occurs at
$s=0$. As a result, we have%
\[
H(r)=P(r,0)=\ln\left(  \frac{1+\nu}{2}\right)  .
\]
Since $\nu=2\bar{n}+1$, we can write the QHB in terms of the mean number of
thermal photons, i.e.,%
\begin{equation}
H(r)=\ln(\bar{n}+1). \label{QHB_the}%
\end{equation}
This is the optimal error exponent for the asymptotic probability of false
negatives, i.e., of confusing a thermal state with the vacuum state.

Let us now consider the thermal state to be the null hypothesis ($\nu^{0}%
:=\nu>1$) while the vacuum state is the alternative hypothesis ($\nu^{1}=1$).
In this case, we derive%
\begin{equation}
P(r,s)=\frac{s}{1-s}\left[  \ln\left(  \frac{1+\nu}{2}\right)  -r\right]  ,
\end{equation}
which leads to the following expression for the QHB%

\begin{equation}
H(r)=\left\{
\begin{array}
[c]{c}%
0\text{~~~~~~~for~}r\geq\ln\left(  \frac{1+\nu}{2}\right)  ~,\\
\\
+\infty\text{~~~~for~}r<\ln\left(  \frac{1+\nu}{2}\right)  ~.~
\end{array}
\right.  \label{H_sing}%
\end{equation}
This is\ related to the minimum probability of confusing the vacuum state with
a thermal state. Note that this is very different from Eq.~(\ref{QHB_the}).

\section{Discrimination of two-mode Gaussian states}

In this section we consider two important classes of two-mode Gaussian states.
The first is the class of Einstein-Podolsky-Rosen (EPR) states, also known as
two-mode squeezed vacuum states. The second (broader)\ class is that of
two-mode squeezed thermal (ST) states, for which the computation of the QHB is numerical.

\subsection{Asymmetric testing of EPR\ correlations}

The expression of the QHB in the case of EPR states is easy to derive. Since
EPR states are pure, the QHB $H(r)$ is given by $H_{F}(r)$ of Eq.~(\ref{QHBF2}%
). As a result, we need only to compute the fidelity between the two states.

An EPR state has zero mean and CM%
\begin{equation}
\mathbf{V}_{\text{EPR}}(\mu)=\left(
\begin{array}
[c]{cc}%
\mu\mathbf{I} & \sqrt{\mu^{2}-1}\mathbf{Z}\\
\sqrt{\mu^{2}-1}\mathbf{Z} & \mu\mathbf{I}%
\end{array}
\right)  , \label{Vepr}%
\end{equation}
with $\mu\geq1$, $\mathbf{I}$ is the $2\times2$ identity matrix and%
\begin{equation}
\mathbf{Z}:=\left(
\begin{array}
[c]{cc}%
1 & 0\\
0 & -1
\end{array}
\right)  .
\end{equation}
Given two EPR states with parameters $\mu_{0}$ and $\mu_{1}$, their fidelity
is computed via Eq.~(\ref{Fid_FormulaNEW}), yielding%
\begin{equation}
F=\frac{4}{\sqrt{\det\mathbf{L}}},
\end{equation}
where $\mathbf{L}=\mathbf{V}_{\text{EPR}}(\mu_{0})+\mathbf{V}_{\text{EPR}}%
(\mu_{1})$. After simple algebra, we find%
\begin{equation}
F=\frac{2}{1+\mu_{0}\mu_{1}-\sqrt{(\mu_{0}^{2}-1)(\mu_{1}^{2}-1)}},
\end{equation}
to be used in Eq.~(\ref{QHBF2}).

\subsection{Squeezed thermal states}

In this section we consider symmetric ST states $\rho(\mu,c)$, which are
Gaussian states with zero mean and CM%
\begin{equation}
\mathbf{V}_{\text{ST}}(\mu,c)=\left(
\begin{array}
[c]{cc}%
\mu\mathbf{I} & c\mathbf{Z}\\
c\mathbf{Z} & \mu\mathbf{I}%
\end{array}
\right)  , \label{symCM}%
\end{equation}
where $\mu\geq1$ and $\left\vert c\right\vert \leq\mu$~\cite{Bfide,Bfide2} (in
particular, without loss of generality, we can assume $c\geq0$). These are
called symmetric because they are invariant under permutation of the two
modes~\cite{asy}.

Note that, for $c=0$, we have no correlations, and the ST state is a
tensor-product of thermal states, i.e., $\rho(\mu,0)=\rho_{\text{th}}%
(\mu)^{\otimes2}$. For $c=\sqrt{\mu^{2}-1}$ the correlations are maximal, and
the ST state becomes an EPR state, i.e., $\rho(\mu,\sqrt{\mu^{2}-1}%
)=\rho_{\text{EPR}}(\mu)$. Finally, for $c=\mu-1$, we have maximal separable
correlations. In other words, $\rho(\mu,\mu-1)$ is the separable ST state with
the strongest correlations (e.g., highest discord).

The symplectic decomposition of a symmetric ST\ state is known. From the CM of
Eq.~(\ref{symCM}), one can check that the symplectic spectrum is degenerate
and given by the single eigenvalue
\begin{equation}
\nu=\sqrt{\mu^{2}-c^{2}}. \label{degEIG}%
\end{equation}
The symplectic matrix $\mathbf{S}$ which diagonalizes $\mathbf{V}_{\text{ST}%
}(\mu,c)$ in Williamson form $\nu(\mathbf{I}\oplus\mathbf{I})$ is given by%
\begin{equation}
\mathbf{S}=\left(
\begin{array}
[c]{cc}%
\omega_{+}\mathbf{I} & \omega_{-}\mathbf{Z}\\
\omega_{-}\mathbf{Z} & \omega_{+}\mathbf{I}%
\end{array}
\right)  , \label{diagS}%
\end{equation}
where%
\begin{equation}
\omega_{\pm}:=\sqrt{\frac{\mu\pm\nu}{2\nu}}. \label{omeDEF}%
\end{equation}

As a result, the s-overlap between two symmetric ST states, $\rho_{0}$ and
$\rho_{1}$, can be computed using the simplified formulas%
\begin{align}
\Pi_{s}  &  =4~G_{s}^{2}(\nu^{0})~G_{1-s}^{2}(\nu^{1}),\label{stsEQ1}\\
\boldsymbol{\Sigma}_{s}  &  =\Lambda_{s}(\nu^{0})~\mathbf{S}_{0}\mathbf{S}%
_{0}^{T}+\Lambda_{1-s}(\nu^{1})~\mathbf{S}_{1}\mathbf{S}_{1}^{T},
\label{stsEQ2}%
\end{align}
where $\nu^{0}$ ($\nu^{1}$) is the degenerate eigenvalue of $\rho_{0}$
($\rho_{1}$), computed according to Eq.~(\ref{degEIG}), and $\mathbf{S}_{0}$
($\mathbf{S}_{1}$) is the corresponding diagonalizing symplectic matrix,
computed according to Eqs.~(\ref{diagS}) and~(\ref{omeDEF}).

\begin{figure}[ptbh] \vspace{-1.2cm}
\par
\begin{center}
\includegraphics[width=0.50\textwidth] {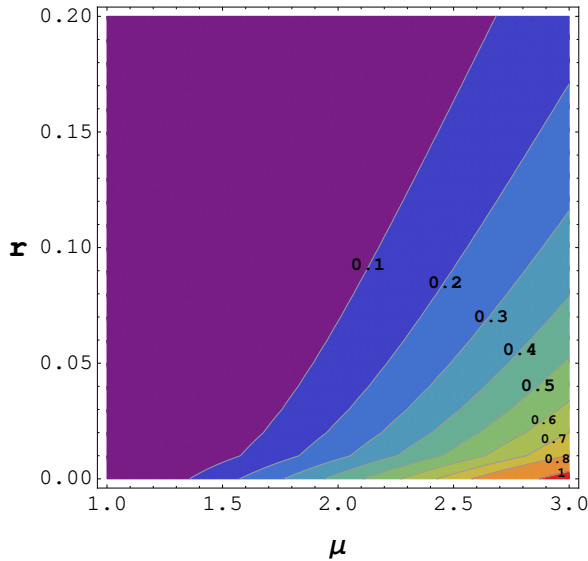}
\end{center}
\par
\vspace{-0.75cm}\caption{(Color online). Asymmetric discrimination
between the thermal state $\rho_{0}=\rho(\mu,0)$ and the ST state
$\rho_{1}=\rho(\mu ,\mu-1)$ with maximal separable correlations.
We plot the QHB as a function of the thermal variance $\mu$ and
the false-positive parameter $r$. As we can see
the QHB\ improves for lower $r$ and for higher $\mu$.}%
\label{Pic}%
\end{figure}

Let us start with simple cases involving the asymmetric testing of
correlations with specific ST states. First we consider the asymmetric
discrimination between the uncorrelated thermal state $\rho_{0}=\rho(\mu,0)$
as null hypothesis and the correlated (but separable) ST state $\rho_{1}%
=\rho(\mu,\mu-1)$ as alternative hypothesis. A false negative
corresponds to concluding that there are no correlations where
they are actually present~\cite{Other}. It is straightforward to
derive their degenerate symplectic eigenvalues which are simply
$\nu^{0}=\mu$ and $\nu^{1}=\sqrt {2\mu-1}$. Then, we have
$\mathbf{S}_{0}=\mathbf{I}\oplus\mathbf{I}$, while
$\mathbf{S}_{1}$ can be easily computed from Eqs.~(\ref{diagS})
and~(\ref{omeDEF}). By substituting these into Eqs.~(\ref{stsEQ1})
and~(\ref{stsEQ2}), we can compute the s-overlap
$C_{s}=\Pi_{s}/\sqrt {\det\boldsymbol{\Sigma}_{s}}$ and therefore
the QHB $H(r)$ via Eq.~(\ref{QHB2}). The results are plotted in
Fig.~\ref{Pic}, for values of thermal variance $\mu$ up to $3$
(i.e., from zero to $1$ mean photon) and small values of the
parameter $r$, bounding the rate of false-positives. As expected,
the QHB\ improves for decreasing $r$ and increasing $\mu
$.

\begin{figure}[ptbh] \vspace{-0.9cm}
\par
\begin{center}
\includegraphics[width=0.48\textwidth] {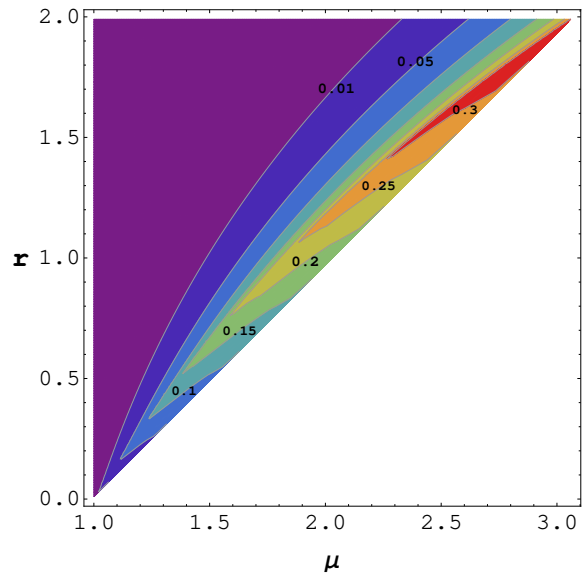}
\end{center}
\par
\vspace{-0.6cm}\caption{(Color online). Asymmetric discrimination between the
thermal state $\rho_{0}=\rho(\mu,0)$ and the EPR state $\rho_{1}%
=\rho_{\text{EPR}}(\mu)$. We plot the QHB as a function of the thermal
variance $\mu$ and the false-positive parameter $r$. The QHB\ improves for
lower $r$ and for higher $\mu$. In particular, there is a threshold value
after which the QHB becomes infinite (white region).}%
\label{Pic2}%
\end{figure}

Now let us consider the asymmetric discrimination between $\rho_{0}=\rho
(\mu,0)$ and the EPR state $\rho_{1}=\rho_{\text{EPR}}(\mu)$, i.e., the most
correlated and entangled ST state~\cite{Other}. Thanks to the simple
symplectic decomposition of the EPR\ state ($\nu^{1}=1$), we can further
simplify the previous Eqs.~(\ref{stsEQ1})-(\ref{stsEQ2}) and write%
\begin{equation}
\Pi_{s}=4~G_{s}^{2}(\mu),~\boldsymbol{\Sigma}_{s}=\Lambda_{s}(\mu
)~(\mathbf{I}\oplus\mathbf{I})+\mathbf{V}_{\text{EPR}}(\mu),
\end{equation}
with $\mathbf{V}_{\text{EPR}}(\mu)$ being given by
Eq.~(\ref{Vepr}). As before, we compute the QHB which is plotted
in Fig.~\ref{Pic2}, for $1\leq \mu\leq3$ and $r\leq2$. As expected
the QHB\ improves for decreasing $r$ and increasing $\mu$. Note a
discontinuity identifying two regions, one where the QHB is
finite, and the other where it is infinite (white region in the
figure).

In fact, by expanding the term $P(r,s)$ in Eq.~(\ref{QHB2}) for $s\rightarrow
1^{-}$, that we find
\begin{equation}
P(r,s)\simeq\frac{N}{s-1}+O(s-1),
\end{equation}
where%
\begin{equation}
N:=r-\ln\left(  \frac{1+3\mu^{2}}{4}\right)  .
\end{equation}
For values of $r$ and $\mu$ such that $N>0$, we find that the term $P(r,s)$
diverges at the border, making the QHB\ infinite. For a given $r$, this
happens when
\begin{equation}
\mu>\tilde{\mu}(r):=\sqrt{\frac{4e^{r}-1}{3}}.
\end{equation}

\begin{figure}[tbh] \vspace{+0.1cm}
\par
\begin{center}
\includegraphics[width=0.48\textwidth] {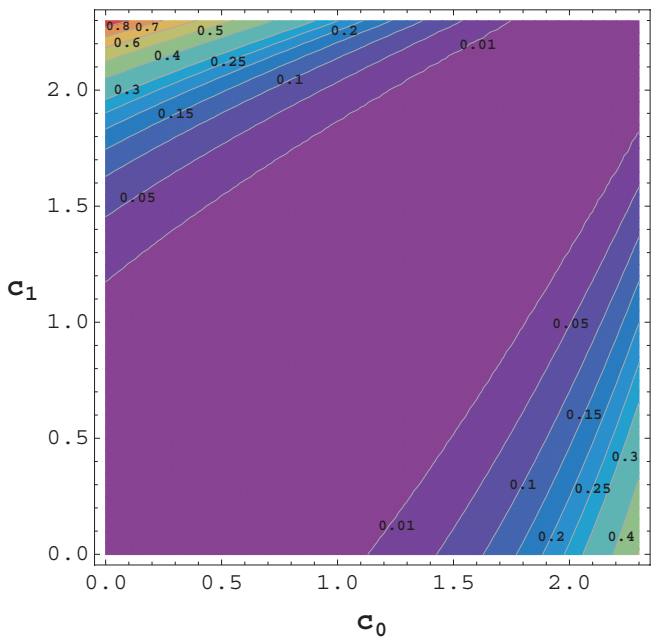}
\end{center}
\par
\vspace{-0.6cm}\caption{(Color online). Asymmetric discrimination between two
ST states with the same thermal variance ($\mu=3$) but different correlations
$c_{0}$ and $c_{1}$. Setting $r=0.1$, we plot the QHB as a function of $c_{0}$
and $c_{1}$. We can see that the QHB increases orthogonally to the bisector
$c_{0}=c_{1}$.}%
\label{Pic3}%
\end{figure}

Finally, we consider the most general scenario in the asymmetric
testing of correlations with ST states. In fact, we consider two
generic ST states, $\rho(\mu,c_{0})$ and $\rho(\mu,c_{1})$, with
the same thermal noise but differing amounts of correlation. For
this computation, we use Eqs.~(\ref{degEIG})-(\ref{omeDEF}) with
$c=c_{0}$ or $c_{1}$, to be replaced in
Eqs.~(\ref{stsEQ1})-(\ref{stsEQ2}), therefore deriving the
s-overlap and the QHB. At small thermal variance ($\mu=3$) and for
the numerical value $r=0.1$, we plot the QHB\ as a function of the
correlation parameters $c_{0}$ and $c_{1}$. As we can see from
Fig.~\ref{Pic3}, the QHB is not symmetric with respect to the
bisector $c_{0}=c_{1}$ (where it is zero) and increases away from
this line.

\section{Conclusion}

In this work we have considered the problem of asymmetric quantum hypothesis
testing by adopting the recently-developed tool of the quantum Hoeffding bound
(QHB). After a brief review of these notions, we have shown how the QHB can be
simplified in some cases (pure states) and estimated using other
easier-to-compute bounds based on simple algebraic inequalities.

In particular, we have applied the theory of asymmetric testing to multimode
Gaussian states, providing a general recipe for the computation of the QHB in
the Gaussian setting. Using this recipe, we have found analytic formulas and
shown numerical results for important classes of one-mode and two-mode
Gaussian states. In particular, we have studied the behavior of the QHB in the
low energy regime, i.e., considering Gaussian states with a small average
number of photons.

Our results could be exploited in protocols of quantum information with
continuous variables. In particular, they could be useful for reformulating
Gaussian schemes of quantum state discrimination and quantum channel
discrimination in such a way as to give more importance to one of the quantum
hypotheses. This asymmetric approach could be the most suitable in the
development of quantum technology for medical applications.

\section*{Acknowledgments}

This work has been supported by EPSRC (EP/J00796X/1). G.S. has been supported
by an EPSRC DTA grant. The authors thank C. Ottaviani and S. Pirandola for
enlightening discussions.


\begin{thebibliography}{99}
\bibitem {NoteGG}We can write $H(r)=\max\left\{  P(r,0),\sup_{0<s<1}%
P(r,s)\right\}  $, where $P(r,0)=-\ln C_{0}=0$ can be neglected and
\[
\sup_{0<s<1}P(r,s)=\sup_{0<s<1}\frac{-r~s-\ln F}{1-s}=\left\{
\begin{array}
[c]{c}%
\ln\frac{1}{F}\text{~~for~}r\geq\ln\frac{1}{F},\\
\\
+\infty\text{~~for~}r<\ln\frac{1}{F}.
\end{array}
\right.
\]
\bibitem {SPEDcomm}More generally, for any two vectorial operators
$\mathbf{a}$ and $\mathbf{b}$, we can express their commutation relations in
the compact form $[\mathbf{a},\mathbf{b}^{T}]:=\mathbf{ab}^{T}-(\mathbf{ba}%
^{T})^{T}$.
\bibitem {asy}Extension to asymmetric ST states\ is only technical.
\bibitem {Other}For brevity we do not consider the other case where the ST
state is the null hypothesis and the thermal state is the alternative
hypothesis. This case is included in the our final analysis for generic ST states.
\end{thebibliography}
\end{document}